\documentclass[twocolumn,aps,pra,superscriptaddress,showpacs,longbibliography]{revtex4-1} 
\usepackage[english]{babel}               % Voor nederlandstalige hyphenatie (woordsplitsing)
\usepackage{amsmath}                    % Uitgebreide wiskundige mogelijkheden
\usepackage{amssymb}                    % Voor speciale symbolen zoals de verzameling Z, R...
\usepackage{makeidx}                    % Om een index te maken
\usepackage{url}                        % Om url's te verwerken
\usepackage{graphicx}                   % Om figuren te kunnen verwerken
\usepackage[small,bf,hang]{caption}     % Om de captions wat te verbeteren
\usepackage{xspace}                     % Magische spaties na een commando
\usepackage[latin1]{inputenc}           % \textsl{}Om niet ascii karakters rechtstreeks te kunnen typen
\usepackage{float}                      % Om nieuwe float environments aan te maken. Ook optie H!
\usepackage{flafter}                    % Opdat floats niet zouden voorsteken
\usepackage{listings}                   % Voor het weergeven van letterlijke text en codelistings
\usepackage[version=3]{mhchem}          % Voor elegante scheikundige formules
\usepackage{subcaption}
\usepackage{bm}
\usepackage[a4paper,plainpages=false,pdfstartview={XYZ null null 0.75}, colorlinks=true, citecolor=blue,linkcolor=PineGreen]{hyperref}    % Om hyperlinks te hebben in het pdfdocument.
\usepackage[usenames,dvipsnames]{color}
\usepackage{cleveref}
\usepackage[colorinlistoftodos,prependcaption,textsize=tiny]{todonotes}
\usepackage{tikz}
\makeindex                              % Om een index te genereren.

\newcommand{\diff}{\ensuremath{\mathrm{d}}}

\newcommand{\partieel}[2]{\ensuremath{\frac{\partial {#1}}{\partial {#2}}}}

\newcommand{\vt}[1]{\ensuremath{\boldsymbol{#1}}} % vector in juiste lettertype
\newcommand{\kmax}{\ensuremath{k_{\text{max}}}}
\begin{document}
\title{Investigation into spiral phase plate contrast in optical and electron microscopy}
\author{Roeland Juchtmans}
\affiliation{EMAT, University of Antwerp, Groenenborgerlaan 171, 2020 Antwerp, Belgium}
\author{Laura Clark}
\affiliation{EMAT, University of Antwerp, Groenenborgerlaan 171, 2020 Antwerp, Belgium}
\author{Axel Lubk}
\affiliation{Triebenberg Laboratory, Institute of Structure Physics, Technische Universit\"at Dresden, 01062 Dresden, Germany}
\author{Jo Verbeeck}
\affiliation{EMAT, University of Antwerp, Groenenborgerlaan 171, 2020 Antwerp, Belgium}
\begin{abstract}
    The use of phase plates in the back focal plane of a microscope is a well established technique in optical microscopy to increase the contrast of weakly interacting samples and is gaining interest in electron microscopy as well. In this paper we study the spiral phase plate (SPP), also called helical, vortex, or two-dimensional Hilbert phase plate, that adds an angularly dependent phase of the form $e^{i\ell\phi_k}$ to the exit wave in Fourier space. In the limit of large collection angles, we analytically calculate that the average of a pair of $\ell=\pm1$ SPP filtered images is directly proportional to the gradient squared of the exit wave, explaining the edge contrast previously seen in optical SPP work. We discuss the difference between a clockwise--anticlockwise pair of SPP filtered images and derive conditions under which the modulus of the wave's gradient can be seen directly from one SPP filtered image.
    %Finally, we demonstrate how with three images, one without and one with each of an $\ell=\pm1$ SPP, may give enough information to reconstruct both the amplitude and the phase of the exit wave.
    This work provides the theoretical background to interpret images obtained with a SPP, thereby opening new perspectives for new experiments to study for example magnetic materials in an electron microscope.
\end{abstract}
\maketitle
\section{Introduction}
Inserting phase plates in the back focal plane of a microscope, whether optical \cite{Born1999,Martin1966,Barer1953} or electron \cite{Danev2001,Danev2008,Kaneko2006}, is a well established technique to increase the contrast of transparent objects. This way even pure phase objects can be made visible. 
In optical microscopy this technique was first developed with the introduction of the Zernike phase plate \cite{Zernike1942,Zernike1942a} adding a $\pm\pi/2$ phase difference to the scattered part of the wave compared to the transmitted part and was followed by other types of phase-imaging methods including the differential interference contrast (DIC) microscopy \cite{Pluta1994} and Hoffman modulation contrast (HMC) microscopy \cite{Hoffman1975}.
In electron microscopy charging and contamination impedes the design of workable phase plates significantly.
A first equivalent to the Zernike phase plate, the Boersch phase plate \cite{Boersch1947}, adds an extra phase to the scattered wave using a charged metallic ring in the Fourier plane \cite{Matsumoto1996,Schultheiss2006,Majorovits2007}. An alternative way of imparting a phase shift on the scattered beam is to use a thin film carbon sheet with a hole in the center \cite{Danev2001,Shimada2007,Yamaguchi2008}. Both techniques have been demonstrated experimentally, but show complications that prevent general application. Only recently a workable electron Zernike phase plate called the Volta phase plate was designed where a thin film is introduced in the back focal plane. The unscattered beam modifies the surface of the film, giving the central beam its phase shift \cite{Danev2014} and has been successfully applied in cryo-tomography \cite{Asano2015,Mahamid2016} and single particle analysis \cite{Khoshouei2016}.
Also an electron equivalent to the DIC method has been shown \cite{Danev2002, Kaneko2005,Kaneko2006,Nitta2009}.
Depending on the shape of the phase plate, the transparent object is made visible in different ways. 
Images made with a Zernike phase plate are (to a first-order approximation) direct phase contrast images, where the intensity in the image is proportional to the phase shift of the wave.
In images made with the DIC and HMC method, the transparent objects are revealed as if they are illuminated from one side, appearing bright on one side and casting a shadow on the other.

Some particularly interesting types of phase plate are the one- and two-dimensional Hilbert phase plates \cite{Davis2000,Danev2002}.
In the one-dimensional case, one side of the wave in Fourier space is given a $\pi$ phase shift with respect to the other.
Edges perpendicular to this direction then appear as bright lines in the image.
In order to remove the directional dependency of the edge-enhancement, the two-dimensional radially symmetric Hilbert phase plate can be used that adds a phase of the form $\exp(i\ell\phi_k)$ to the wave in Fourier space, where $\phi_k$ is the angular coordinate with respect to the center of the beam and $\ell$ is an integer number. When $\ell=\pm1$, there is a $\pi$-phase shift across any diameter of the phase plate.
Note that these phase plates, called vortex, helical or spiral phase plates (SPP), can also be used to generate vortex beams that are characterized by a wavefunction of the form $\Psi(\vt{r})=\psi(r)e^{i\ell\phi}e^{ik_zz}$ \cite{Nye1974,Allen1992}, which have their own fields of application in optics \cite{Galajda2001,Friese,Luo,He,Swartzlander2007,Serabyn2010,Berkhout2008,wangterabit,vortexnot} and electron microscopy \cite{Verbeeck2013,Juchtmans2015}.
In phase-contrast microscopy, these phase plates have been proven to be useful in optics \cite{Furhapter2005}, but also in electron microscopy they are attracting increasingly more attention as an edge--enhancement technique \cite{Blackburn2014} or to detect the chirality of crystals \cite{Juchtmans2015a}. 
For charged particles the one-dimensional Hilbert phase shift can be induced using the Aharonov-Bohm phase shift of a thin magnetized wire placed along the diameter of an aperture \cite{Nagayama2008}.
In a same way, the angularly dependent phase of the two-dimensional Hilbert phase can be generated near the tip of a magnetized needle \cite{Beche2013,Blackburn2014,Beche2016a}.
Although alternative SPPs for electrons have been investigated, e.g., fork gratings \cite{Verbeeck2010} or helically shaped lenses made of a transparent material \cite{Beche2016}, the magnetic phase plates have the advantage of only blocking a relatively small part of the aperture in the objective plane, such that a maximal amount of scattered electrons can contribute to the signal.

In previous work \cite{Juchtmans2015b}, we related the intensity of images obtained with a SPP to the local OAM decomposition of the exit wave. However, in ref. \cite{Davis2000} it was shown that the same phase plate enhances the contrast at edges of transparent objects in all directions, and in ref. \cite{Furhapter2005} it was suggested that the intensity in the images is related to the gradient of the exit wave.

In this paper, we analytically investigate what a SPP with $\ell=\pm1$ reveals about the exit wave and its partial derivatives within the approximation of large collection angles and verify this with numerical simulations.
Based on this, we discuss under which circumstances the SPP filtered images can be directly linked to the gradient of the wave, and when this approximation no longer holds.
Finally, we shortly discuss how the combined information of three images made with a $\ell=-1$, $\ell=0$ and a $\ell=1$ SPP may provide enough information to make a full exit wave reconstruction of both the phase and intensity.

\section{Images with $\ell=\pm 1$ spiral phase plates}
%\begin{figure}[t]
%	\includegraphics[width=\columnwidth]{T1rk}
%	\caption{The radial profile, $T^{\kmax}(r)$, of the Fourier transform for SPP apertures with different radii  $\kmax\in\{1,2,3,4\}\text{nm}^-$.\label{fig:Tkmax}}
%\end{figure}

%\begin{center}
%	\begin{figure}[t]
%		\begingroup%
%		\setlength{\unitlength}{\columnwidth}%
%		\makeatother
%		\begin{picture}(1,0.615)%
%		\put(0,0){\includegraphics[width=\unitlength]{fig1}}%
%		\put(.76,0.495){\color[rgb]{0,0,0}\makebox(0,0)[lb]{\smash{$\kmax=1/$\AA}}}%
%		\put(.76,0.440){\color[rgb]{0,0,0}\makebox(0,0)[lb]{\smash{$\kmax=2/$\AA}}}%
%		\put(.76,0.385){\color[rgb]{0,0,0}\makebox(0,0)[lb]{\smash{$\kmax=3/$\AA}}}%
%		\put(.76,0.330){\color[rgb]{0,0,0}\makebox(0,0)[lb]{\smash{$\kmax=4/$\AA}}}%
%		\put(.85,0.03){\color[rgb]{0,0,0}\makebox(0,0)[lb]{\smash{$r$ [\AA]}}}%
%		\put(.216,0.03){\color[rgb]{0,0,0}\makebox(0,0)[lb]{\smash{2}}}%
%		\put(.404,0.03){\color[rgb]{0,0,0}\makebox(0,0)[lb]{\smash{4}}}%
%		\put(.592,0.03){\color[rgb]{0,0,0}\makebox(0,0)[lb]{\smash{6}}}%
%		\put(.78,0.03){\color[rgb]{0,0,0}\makebox(0,0)[lb]{\smash{8}}}%
%		
%		%\put(-2.6766732,-0.07931681){\color[rgb]{0,0,0}\makebox(0,0)[lt]{\begin{minipage}{3.95687558\unitlength}\raggedright \end{minipage}}}%
%		
%		\end{picture}%
%		\endgroup%
%		\caption{The radial profile, $T^{\kmax}(r)$, of the Fourier transform for SPP apertures with different radii  $\kmax\in\{1,2,3,4\}\text{\AA}^{-1}$.\label{fig:Tkmax}}
%	\end{figure}
%\end{center}

	\begin{figure}[t]
		\centering
		\includegraphics[width=\columnwidth]{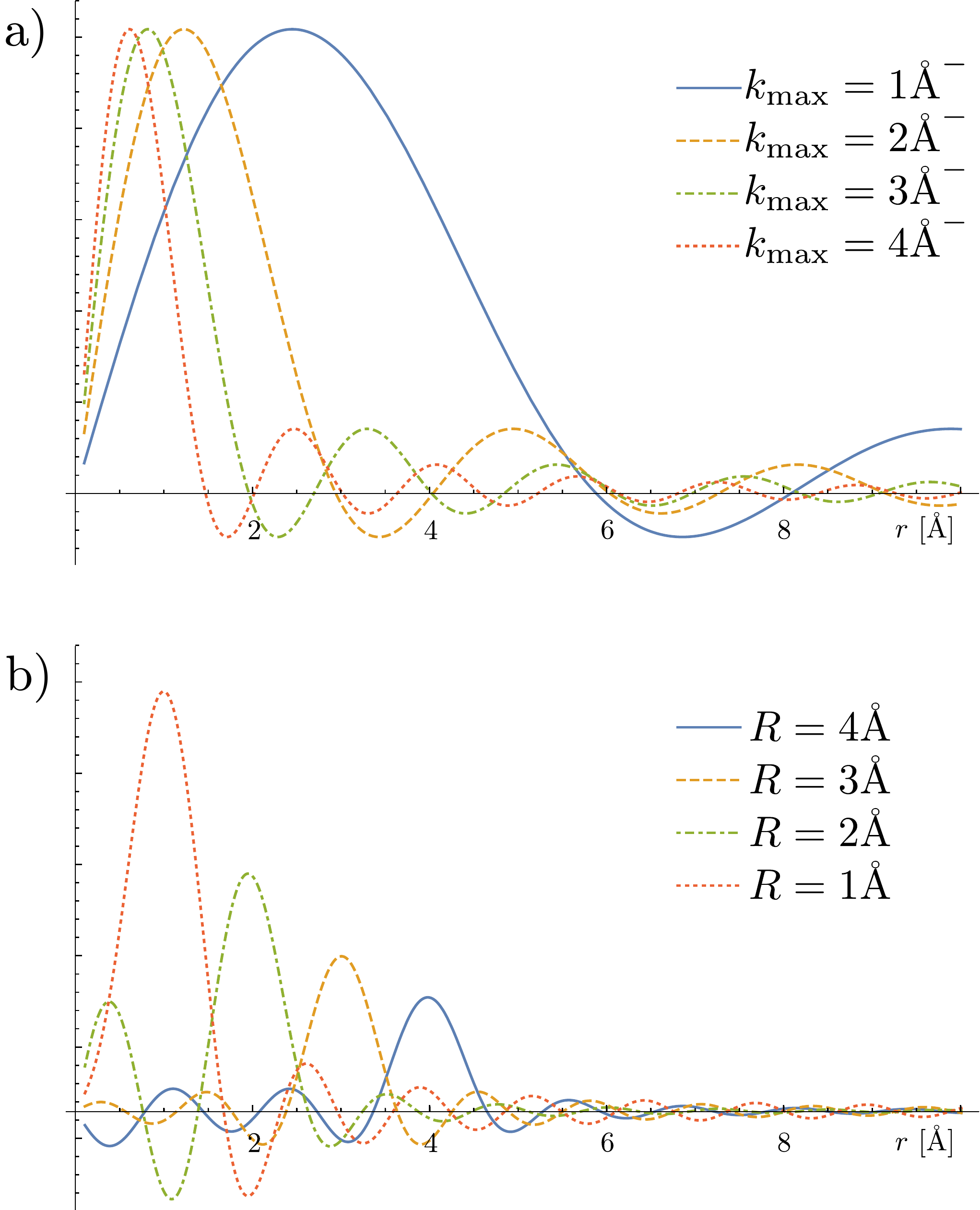}%
		\caption{The radial profiles of the FT of flat amplitude SPP apertures with different radii  $\kmax\in\{1,2,3,4\}\text{\AA}^{-1}$ (a) and for Bessel modulated amplitude SPP apertures with \kmax=5\AA~$^-$ for different $R$ (b)\label{fig:Tkmax}}
	\end{figure}
 In the following $\Psi(\vt{r})=\Psi(r,\phi)$ and $\tilde{\Psi}(\vt{k})=\tilde{\Psi}(k,\phi_k)$ denote a two-dimensional section of the three-dimensional wave in real and Fourier space respectively.

Let $\Psi(\bm r)$ be a two-dimensional wave of interest, e.g., scalar electromagnetic wave or electron wave after interaction with a sample. Introducing a spiral phase plate (SPP) into the back focal plane of a microscope means that we add a phase of the form $e^{i\ell\phi_k}$, with $\phi_k$ the angular coordinate and $\ell=\pm1$ an integer, to the scattered wave in the diffraction plane. Mathematically this means we multiply the Fourier transform (FT) of the exit wave with an angularly dependent phase factor
\begin{align}
\tilde{\Psi}^{\pm}(\bm k)=\begin{cases}
\tilde{\Psi}(\bm k)e^{\pm i \phi_k}\hspace{1cm} &\bm (k\neq0)\\
0 &(k=0),
\end{cases}
\end{align}
where we set the amplitude of the wave to zero in the origin since the phase factor $e^{\pm i \phi_k}$ is undefined in this point.
To obtain the resulting image, we must propagate to real space by taking the inverse Fourier transform 
\begin{align}
{\Psi}^{\pm}&=\mathcal{F}^{-1}[\tilde{\Psi}(\bm k).e^{ \pm i\phi_k}](\vt{r})\nonumber\\
&=\mathcal{F}^{-1}[\tilde{\Psi}(\bm k)]\otimes \mathcal{F}^{-1}[e^{\pm i\phi_k}]\nonumber\\
&=\Psi(\bm r)\otimes \mathcal{F}^{-1}[e^{\pm i\phi_k}],\label{eq:Convolution}
\end{align}
where we used the convolution theorem in the second transformation. The resulting image therefore is given by the convolution of the exit wave with the Fourier transform of the SPP. It is easy to show that the latter has a radially symmetric amplitude with an angularly dependent phase factor $e^{\pm i\phi}$, i.e.,
\begin{align}
&\mathcal{F}^{-1}[e^{\pm i\phi_k}]=\int_{0}^{\infty}dk\int_{0}^{2\pi}d\phi_k\, ke^{\pm i\phi_k}e^{i\vt{k}\cdot\vt{r}}\nonumber\\
	&=\int_{0}^{\infty}dk\int_{0}^{2\pi}d\phi_k \,ke^{\pm i\phi_k}\sum_mi^{m}J_m(kr)e^{im(\phi-\phi_k)}\nonumber\\
	&= 2\pi\sum_m i^m e^{im\phi}\int_{0}^{\infty}dk\,kJ_m(kr)\delta_{m,\pm 1}\nonumber\\
	&= 2\pi i e^{\pm i\phi}\int_{0}^{\infty}dk\,kJ_1(kr),\label{eq:FTSPP}
\end{align}
where we used the Jacobi--Anger identity
\begin{align}
e^{ikr\cos(\phi-\phi_k)}=\sum_mi^{m}J_m(kr)e^{im(\phi-\phi_k)}.
\end{align}
The above expression for the Fourier transform of the SPP is not normalizable, i.e., is a distribution. In reality however, the upper boundary of the integral in \cref{eq:FTSPP} is given by the radius of the SPP aperture ($\kmax$)
\begin{align}
\mathcal{F}^{-1}[\pm SPP]&=2\pi i e^{\pm i\phi}\int_{0}^{\kmax}dk\,kJ_1(kr)\nonumber\\
&=T^{\kmax}(r)e^{\pm i\phi},
\end{align}
where $T^{\kmax}(r)$ is a function that determines the radial profile of the FT of an $\ell=\pm1$ SPP. The function has a global maximum close to zero, but is zero at $r=0$, which gives the Fourier transform of the SPP the typical vortex beam shape, a bright ring with a dark core. As shown in \cref{fig:Tkmax}a, the larger the radius of the SPP aperture, \kmax, the more sharply $T^{\kmax}(r)$ is peaked near $r=0$. 
The radius of this ring is inversely proportional to the size of the phase plate aperture and can be made arbitrarily small when an idealized microscope, aberration corrected at large collection angles, is assumed.
We should note here that, although the ring can be made arbitrarily small, it will not converge to an infinitely sharp delta-ring, because tails will always be present. In order to remove these, a SPP that modulates the amplitude of the wave with a first-order Bessel function, $J_1(kR)$, can be used \cite{Wei2011,Grillo2014b}, where $R$ is a parameter that determines the radius of the ring. In contrast to the constante amplitude (flat) SPP, in the limit of large \kmax, the profile of the FT of the Bessel amplitude modulated aperture converges to a $\delta$-peak. However, since the Bessel modulated aperture also is not normalizable it needs a cut-off at a certain \kmax, giving rise to a broadening of the $\delta$-peak, as can be seen in \cref{fig:Tkmax}b for \kmax=5~\AA$^-$. In what follows all calculations assume a flat SPP, but can easily be extended for a Bessel amplitude modulated SPP.

\begin{center}
	\begin{figure}[t]
		\includegraphics[width=.8\columnwidth]{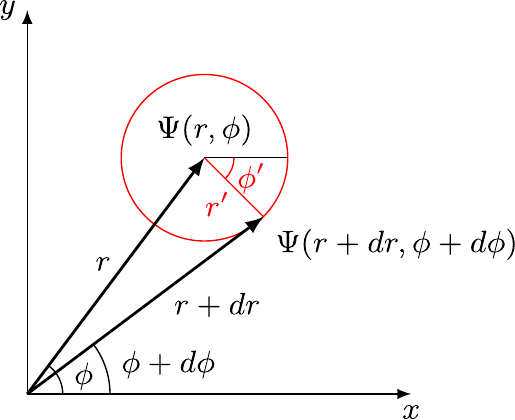}%
		\caption{Scheme of the wavefunction in polar coordinates at points $\bm r$ and $\bm r + \bm{dr}$ given by $\Psi(r,\phi)$ and $\Psi(r+dr,\phi+d\phi)$.\label{ConvCylindrical}}
	\end{figure}
\end{center}

Using a flat SPP, the convolution in \cref{eq:Convolution} can be written as
\begin{align}
\Psi^{\pm}&= \int^{\infty}_{0}dr'\int^{2\pi}_0 d{\phi'}\,r'\Psi(\vt{r}-\vt{r'}) e^{\pm i\phi'}T^{\kmax}(r').\label{eq:Convolution2}
\end{align}
As mentioned above, we can choose $\kmax$ such that $T^{\kmax}(r')$ can be made arbitrarily peaked near $r'=0$. 
For a sufficiently large aperture, $\psi$ can thus be approximated by its first order Taylor expansion in the entire region where $T^{\kmax}(\bm r')$ has a significant weight.
In polar coordinates, defined in \cref{ConvCylindrical}, the first order Taylor expansion of $\psi$ reads
\begin{align}
&\Psi(r+\diff r,\phi+\diff \phi)\approx\Psi(r,\phi)+\nonumber\\
&\partieel{\Psi(r,\phi)}{r}r'\cos(\phi-\phi')+\partieel{\Psi(r,\phi)}{\phi}\frac{r'}{r}\sin(\phi-\phi'), \label{eq:FirstOrder}
\end{align}
and the convolution in \cref{eq:Convolution2} becomes
\begin{align}
\Psi^{\pm}&=\int^{\infty}_{0}dr'\int^{2\pi}_0 d{\phi'}\,r'e^{\pm i\phi'}T^{\kmax}(r')\Big(\Psi(r,\phi)+\nonumber\\
&\left.\partieel{\Psi(r,\phi)}{r}r'\cos(\phi-\phi')+\partieel{\Psi(r,\phi)}{\phi}\frac{r'}{r}\sin(\phi-\phi')\right)\nonumber\\
=&\int^{\infty}_{0}dr'\,r'T^{\kmax}(r')\times\nonumber\\&\left(0-\partieel{\Psi(r,\phi)}{r}r'\pi e^{i\phi}\mp i \partieel{\Psi(r,\phi)}{\phi}\frac{r'}{r}\pi e^{i\phi}\right)\nonumber\\
=&\left(\partieel{\Psi(r,\phi)}{r}\pm \frac{i}{r} \partieel{\Psi(r,\phi)}{\phi}\right)\pi e^{i(\phi+\pi)}\int dr'\,r'^2 T^{\kmax}(r').
\end{align}
The images obtained with an $\ell=+1$ and $\ell=-1$ SPP are given by the modulus squared $\left|\Psi^{\pm}\right|^2$
\begin{align}
&I^+=\nonumber\\&C\left(\partieel{\Psi}{r}\partieel{\Psi^*}{r}- \frac{i}{r}\partieel{\Psi}{r}\partieel{\Psi^*}{\phi}+\frac{i}{r}\partieel{\Psi}{\phi}\partieel{\Psi^*}{r}+\frac{1}{r^2}\partieel{\Psi}{\phi}\partieel{\Psi^*}{\phi}\right)\label{eq:IRight}\\
&I^{-}=\nonumber\\&C\left(\partieel{\Psi}{r}\partieel{\Psi^*}{r}+ \frac{i}{r}\partieel{\Psi}{r}\partieel{\Psi^*}{\phi}-\frac{i}{r}\partieel{\Psi}{\phi}\partieel{\Psi^*}{r}+\frac{1}{r^2}\partieel{\Psi}{\phi}\partieel{\Psi^*}{\phi}\right)\label{eq:ILeft},
\end{align}
with $C=\left|\pi \int^{\infty}_{0}dr'\,r'^2 T^{\kmax}(r')\right|^2$ a normalization factor. From \cref{eq:IRight} and \cref{eq:ILeft} we indeed see that, in the limit of large \kmax, there is a clear relation between the SPP filtered images and the first partial derivative of the exit wave $\Psi$. However, images taken with oppositely handed SPPs, in general, are not equal to each other and therefor can not be both proportional to the gradient of the exit wave as might be expected from the observation of enhanced edge contrast \cite{Davis2000,Furhapter2005}.

\section{Average and difference of $\ell=\pm1$ SPP filtered images}
We can, however, measure the gradient of the wave directly by looking at the average of two opposite SPP filtered images
\begin{align}
\frac{1}{2}\left(I^++I^{-}\right)=&C\left(\partieel{\Psi}{r}\partieel{\Psi^*}{r}+\frac{1}{r^2}\partieel{\Psi}{\phi}\partieel{\Psi^*}{\phi}\right)\nonumber\\
=&C\left|\bm \nabla_\perp \Psi\right|^2,\label{eq:I+}
\end{align}
where $\bm \nabla_\perp$ is the two dimensional gradient of the exit wave in the directions perpendicular to the optical axis. 

Accordingly, individual $\ell=\pm1$ SPP filtered images are proportional to the square of the magnitude of the gradient, only if the difference between opposite SPP images is zero. 
This difference is given by
\begin{align}
\frac{1}{2}\left(I^+-I^{-}\right)=&\frac{iC}{r}\left(\partieel{\Psi}{\phi}\partieel{\Psi^*}{r}-\partieel{\Psi}{r}\partieel{\Psi^*}{\phi}\right)\nonumber\\
&=C'(\vt{\nabla}\times\vt{J})_z,\label{eq:14}
\end{align}
with $\vt{J}$ the probability current density
\begin{align}
\vt{J} &=C''\left(\Psi^*\bm \nabla\Psi-\Psi\bm \nabla \Psi^*\right),
\end{align}
where $C''$ is equal to $i/2$ for photons and $\frac{\hbar}{2mi}$ for electrons.

Using two opposite SPPs thus allows us to directly measure the magnitude of the gradient of the exit wave by adding, and the curl of the current density by subtracting the two images. The latter gives an alternative to the setup using differential astigmatism defoci for measuring the complete probability current, including the solenoidal one \cite{Lubk2015}. Following M. V. Berry \cite{Berry2009}, we call the curl of the current density simply the current vorticity.

In what follows, it will be more convenient to work in Cartesian coordinates
\begin{align}
\frac{1}{2}\left(I^++I^{-}\right)=&C\left(\partieel{\psi}{x}\partieel{\psi^{*}}{x}+\partieel{\psi}{y}\partieel{\psi^{*}}{y}\right) \label{eq:I+Cart}\\
\frac{1}{2}\left(I^+-I^{-}\right)=&iC\left(\partieel{\psi}{x}\partieel{\psi^{*}}{y}-\partieel{\psi}{y}\partieel{\psi^{*}}{x}\right)\label{eq:I-Cart}
\end{align}
To understand the meaning of \cref{eq:I+Cart} and \cref{eq:I-Cart}, we decompose the exit wave into its amplitude and phase
\begin{align}
\Psi(\vt{r})=A(\vt{r})e^{i\varphi(\vt{r})}.
\end{align}
For the average of the two opposite SPP filtered images, we then obtain (omitting the normalization constant $C$),
\begin{align}
&\frac{1}{2}\left(I^++I^{-}\right)\nonumber\\&=\left|\left(\partieel{A}{x}e^{i\varphi}+iAe^{i\varphi}\partieel{\varphi}{x}\right)\vt{e}_x+\left(\partieel{A}{y}e^{i\varphi}+iAe^{i\varphi}\partieel{\varphi}{y}\right)\vt{e}_y\right|^2\nonumber\\
&=\left[\left(\partieel{A}{x}\right)^2+\left(A\partieel{\varphi}{x}\right)^2+\left(\partieel{A}{y}\right)^2+\left(A\partieel{\varphi}{y}\right)^2\right],\label{eq:Iplus}
\end{align}
and for the difference
\begin{align}
&\frac{1}{2}\left(I^+-I^{-}\right)\nonumber\\
&=i\left(\partieel{\psi}{x}\partieel{\psi^{*}}{y}-\partieel{\psi}{y}\partieel{\psi^{*}}{x}\right)\nonumber\\
&=\mathrm{Im}\left[\left(\partieel{A}{x}e^{i\varphi}+iAe^{i\varphi}\partieel{\varphi}{x}\right)\left(\partieel{A}{y}e^{-i\varphi}-iAe^{-i\varphi}\partieel{\varphi}{y}\right)\right]\nonumber\\
&=\mathrm{Im}\left[\partieel{A}{x}\partieel{A}{x}+A^2\partieel{\varphi}{x}\partieel{\varphi}{y}+iA\partieel{A}{y}\partieel{\varphi}{x}-iA\partieel{A}{x}\partieel{\varphi}{y}\right]\nonumber\\
& =2A\left[\partieel{A}{y}\partieel{\varphi}{x}-\partieel{A}{x}\partieel{\varphi}{y}\right]\label{eq:Imin}.
\end{align}
Eq. \ref{eq:Imin} shows that there is only a difference between the $\ell=1$ and $\ell=-1$ SPP filtered images in those points, where both of the following conditions are met simultaneously
\begin{itemize}
    \item Both the 2D gradient of the amplitude and the phase are non-zero.
    \item The 2D gradient of the amplitude and the phase are not parallel to each other.
\end{itemize}
When one of these conditions is not fulfilled, oppositely handed SPP filtered images will be identical.
This is the case, for instance, when a plane wave photon or electron interacts with a pure phase object, where the intensity does not change and its gradient is zero everywhere. The two opposite SPP filtered images will be identical and the magnitude of the two-dimensional gradient squared can be seen directly from a single SPP filtered image. 
The same holds for a weak phase/weak amplitude object \cite{Misell1976}, where both the phase and the amplitude are assumed to be proportional to the thickness of the sample. As a consequence their gradients are parallel and, again, no difference in the opposite SPP filtered images will be observed. 

When studying magnetic samples however, an extra phase is induced of which the gradient is not parallel to that of the electrostatic phase shifts. This creates a difference between the two SPP filtered images that can be directly linked to the magnetization state of thin nano-objects.
Whereas in conventional holographic methods the electrostatic contribution to the phase shift has to be separated from the magnetic by flipping the sample upside down \cite{Tonomura1986}, or applying an external magnetic field \cite{Dunin-Borkowski1998}, a SPP analysis would not require manipulation of the sample. Instead, two images with opposite SPPs have to be recorded. When using the magnetic needle setup from B\'ech\'e et al. \cite{Beche2013,Beche2016a}, this can be done simply by changing the objective aperture or flipping the magnetization direction of the needle with an external magnetic field in the condenser plane. Alternatively, using the fork aperture \cite{Verbeeck2010} with high grating frequency, the two images are separated in the image plane, (i.e., the reciprocal of the grating placed in diffraction plane) and can be recorded simultaneously. 
Similar to the holographic methods, both images need to be recorded under the same conditions and aligned properly in order to compare differences between the two.

When the phase object or weak phase/ weak amplitude approximation no longer holds, the current vorticity does not have to be zero anymore and differences appear in images taken with opposite SPPs. This is demonstrated in the next section with numerical simulations of the exit wave of a plane electron wave passing through a 20nm-thick quartz crystal.

\begin{center}
\begin{figure}[t]
\includegraphics[width=\columnwidth]{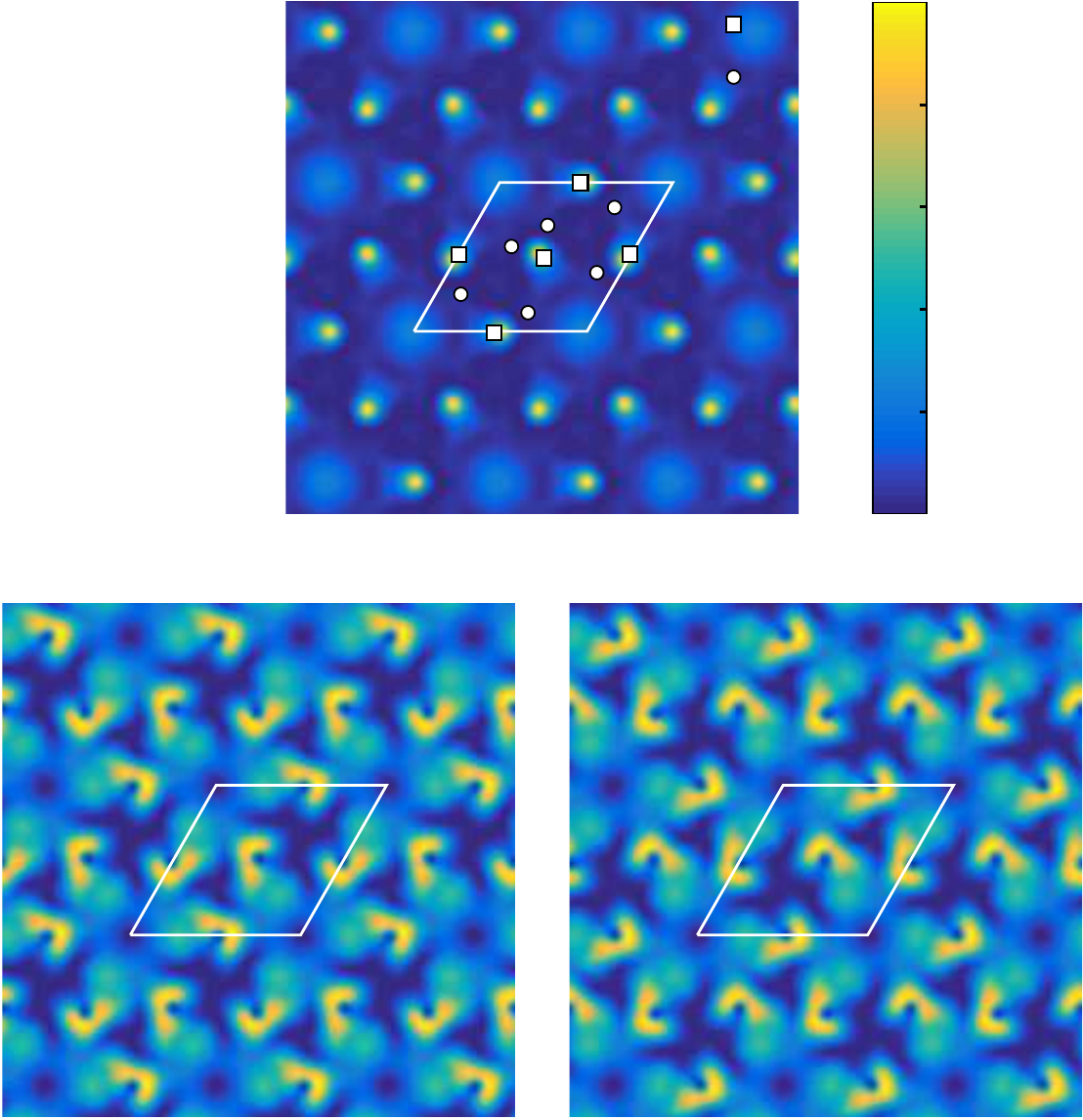}%
\caption{Multislice simulations of TEM images of a 20nm thick quartz crystal (unit cell indicated) without a phase plate(top), with an $\ell=-1$ (left) and with an $\ell=1$ (right) spiral phase plate without intensity modulation (\kmax=5~\AA$^-$). Color map in arbitrary units.\label{fig:TEMImages}}
\end{figure}
\end{center}

\section{Simulation of transmission electron microscopy images\label{sec:Simulation}}
We demonstrate the analytical considerations made in the previous section on a simulated exit wave obtained by multi-slice simulations using the program STEMsim \cite{Rosenauer2007} on a 20nm-thick quartz crystal. Since the simulation gives the amplitude and the phase of the exit wave, quantities such as the gradient of the exit wave or the current vorticity can be computed and compared directly with the SPP filtered images. 

In fig. \ref{fig:TEMImages}, a simulated TEM image is shown together with the images obtained after inserting an $\ell=1$ and $\ell=-1$ SPP. The latter are calculated by taking the amplitude squared of the inverse Fourier transform of the Fourier transform of the exit wave multiplied with a SPP (see \cref{eq:Convolution}). We immediately see that the images obtained with opposite SPP differ significantly, indicating that the intensity in each image can not be linked directly to the gradient of the exit wave.

\begin{center}
	\begin{figure}[t]
		\includegraphics[width=\columnwidth]{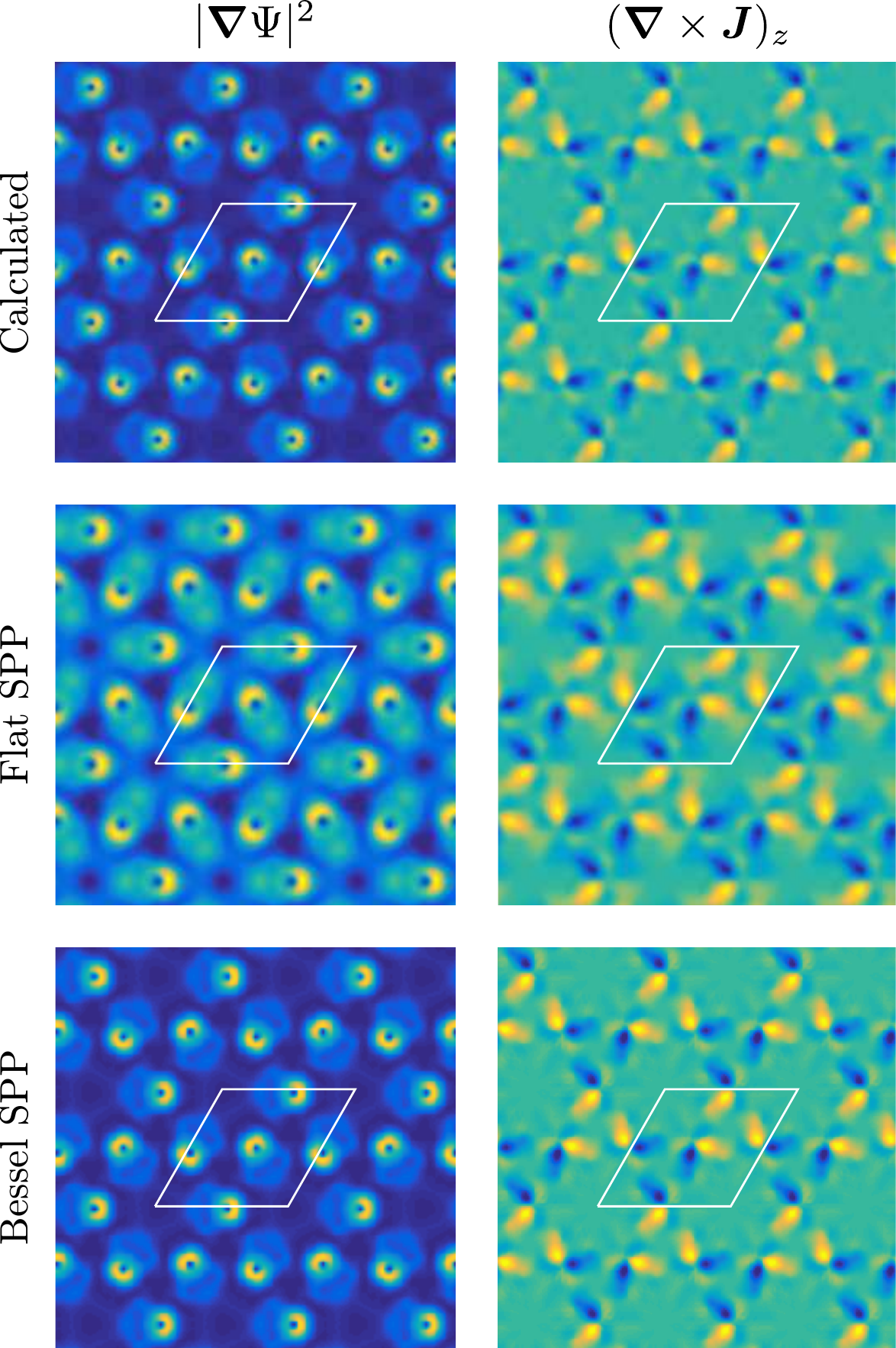}
		\caption{Comparison between $\left|\bm \nabla_\perp \Psi\right|^2$ and $(\bm \nabla \times \bm J)_z$ calculated directly from the exit wave (top) and the average or difference of the SPP filtered images without (middle) and with Bessel modulated amplitude (bottom) (\kmax=5~\AA$^-$, $R=1\AA$).\label{fig:TEMCurlAndGrad}}
	\end{figure}
\end{center}

In fig. \ref{fig:TEMCurlAndGrad} the average and the difference of the SPP filtered images with a flat or Bessel modulated amplitude are compared with the gradient of the exit wave squared, $\left|\bm \nabla_\perp \Psi\right|^2$, and the $z$-component of the curl of the probability current, $(\bm \nabla \times \bm J)_z$, calculated directly from the exit wave. We can see that for both SPPs the average and difference are in good agreement with the calculated gradient and current vorticity images. For the flat SPPs a clear blurring of the images can be seen which is a direct consequence of the tails present in its Fourier transform, see \cref{fig:Tkmax}a. Since in our simulations $\kmax$ is limited by the number of points and the resolution of the image, it can not be made arbitrarily large and these tails will remain to have a significant influence on the images. %Note that this effect appears more present in the summed image than in the subtracted one. The blurring namely is best observed in the regions with low intensity having a lot of details. When looking at the difference however, these details are suppressed by the high intensity differences.
As argued before and shown in \cref{fig:TEMCurlAndGrad}, this blurring effect can be resolved by using a Bessel modulated amplitude SPP. Note that amplitude modulations is far more difficult and generally requires the use of a thin film of transparent material, introducing an increased sensitivity to beam damage and contamination in contrast to flat intensity SPPs, that can be created by magnetic needles or binary fork apertures.

\section{Image wave reconstruction}
Whereas the amplitude of a wave can simply be measured by taking a conventional image, in which the intensity is given by
\begin{align}
I=\left|\psi\right|^2=A^2,
\end{align}
the phase is much harder to measure. However, as shown above, by inserting two oppositely handed SPPs, we get valuable extra information about the wave and the gradient of its amplitude and phase.
 Using $\partieel{I}{x}=2A\partieel{A}{x}$ we can rewrite \cref{eq:Iplus} and \cref{eq:Imin} as
\begin{align}
\begin{cases}
\left[\partieel{I}{y}\partieel{\varphi}{x}-\partieel{I}{x}\partieel{\varphi}{y}\right]&= \frac{1}{2C}(I^+-I^{-})\\
\left(\partieel{\varphi}{x}\right)^2+\left(\partieel{\varphi}{y}\right)^2&= \frac{I^++I^{-}}{2CI}-\left(\frac{1}{A}\partieel{A}{x}\right)^2-\left(\frac{1}{A}\partieel{A}{y}\right)^2\\
&=\frac{I^++I^{-}}{2CI}-\frac{1}{4I^2}\left(\left(\partieel{I}{x}\right)^2+\left(\partieel{I}{y}\right)^2\right),\label{eq:25}
\end{cases}
\end{align}
where $I$ is measured by taking a normal image without a SPP.
Equation \ref{eq:25} represents a set of two first order partial differential equations that may be solved using the method of characteristics, thereby retrieving the phase $\varphi$.
%Knowing the partial derivatives of the phase at each point, in principle the phase can be calculated from integration.
 Note, however, that such a procedure is not straight forward. Problems may arise at crossing characteristics or at amplitude zeros, where the right hand side of the second equation diverges. A detailed elaboration on image wave reconstruction, lies beyond the scope of this paper.

\section{Conclusion}
In order to increase the contrast in optical and electron microscopy images, over the past decades several phase contrast techniques have been developed. In this work we study the effect of a spiral phase plate (SPP) that adds an angularly dependent phase to the wave in the Fourier plane of the form $e^{i\ell\phi}$.
We analytically calculate the effect of an $\ell=\pm1$ SPP in the limit for large $\kmax$ and find that the average of two opposite SPP filtered images is equal to the square of the 2D gradient of the wave and that the difference is proportional to its current vorticity. The latter disappears when the 2D amplitude- and phase-gradient are parallel or if one of them is zero.
We verified these analytical calculations on a simulated exit wave of a plane wave electron passing through a 20nm thick quartz crystal (\cref{fig:TEMCurlAndGrad}).
Our calculations confirm the suggestion made by F\"urhapter et al. \cite{Furhapter2005} that for pure phase objects individual SPP filtered images are proportional to the gradient squared of the exit wave, but also show a further use for the spiral phase plate technique.

We demonstrated how an analysis of two opposite SPP filtered images enables detection of solenoidal currents, such as those occurring in combination with elastic scattering on magnetic samples or chiral inelastic excitations. 
Moreover, we indicate how the combination of a conventional image, an image with an $\ell=-1$ and an $\ell=1$ SPP might give enough information to reconstruct the entire exit wave, thereby opening new pathways to phase retrieval.

\section{acknowledgments}
The authors acknowledge support from the FWO (Aspirant Fonds Wetenschappelijk Onderzoek - Vlaanderen) and the EU under the Seventh Framework Program (FP7) under a contract for an Integrated Infrastructure Initiative, Reference No. 312483-ESTEEM2 and ERC Starting Grant 278510 VORTEX.

\end{document}